\documentclass[aps,prb,reprint,showpacs,floatfix]{revtex4-1}  
\usepackage{amsmath,amssymb,xcolor}
\usepackage[colorlinks,bookmarks=false,citecolor=red,linkcolor=blue,urlcolor=blue]{hyperref}
\usepackage{graphicx,epstopdf}
\usepackage{hyperref}
\hypersetup{
    colorlinks,
    citecolor=blue,
    filecolor=blue,
    linkcolor=blue,
    urlcolor=blue} 
\usepackage{array,multirow}
\newcolumntype{C}[1]{>{\centering\arraybackslash}m{#1}}
\begin{document}
\title{Quantum phases of ferromagnetically coupled dimers on Shastry-Sutherland lattice}
\author{Monalisa Chatterjee}
\email{c.monalisa1993@bose.res.in}
\author{Santanu Pal}
\email{santanu1720@bose.res.in}
\author{Manoranjan Kumar}
\email{manoranjan.kumar@bose.res.in}
\affiliation{S. N. Bose National Centre for Basic Sciences, Block-JD, Sector-III, Salt Lake, Kolkata 700106, India}
\begin{abstract}
The ground state (gs) of antiferromagnetically coupled dimers on the Shastry-Sutherland lattice (SSL)  stabilizes
        many exotic phases and has been extensively studied. The gs properties of ferromagnetically coupled dimers on SSL
        are equally important but unexplored. In this model the exchange coupling along the $x$-axis
        ($J_x$) and $y$-axis ($J_y$) are ferromagnetic and the diagonal exchange coupling ($J$) is antiferromagnetic.
        In this work we explore the quantum phase diagram of
        ferromagnetically coupled dimer model numerically using density matrix renormalization group (DMRG) method.
        We note that in $J_x$-$J_y$ parameter space this model exhibits six interesting phases:(I) stripe $(0,\pi)$,
(II) stripe $(\pi,0)$, (III) perfect dimer, (IV) $X$-spiral, (V) $Y$-spiral and (VI) ferromagnetic phase. Phase boundaries
        of these quantum phases are determined using the correlation functions and gs energies. We also notice
        the correlation length in this system is less than four lattice units in most of the parameter regimes.
        The non-collinear behaviour in $X$-spiral and $Y$-spiral phase and  the dependence of pitch angles on model  parameters are also studied.
\end{abstract}
\maketitle
\section{Introduction}

Frustrated magnetic systems are promising materials to explore exotic phases like 
 dimer \cite{white1994resonating}, non-collinear spin wave \cite{kumar2015level}, spin-liquid \cite{zhou2017quantum,savary2016quantum}, 
non-trivial topological phase \cite{balents2010spin,lee2008end,han2012fractionalized,fu2015evidence} in the ground state (gs). 
In the last couple of decades these systems are extensively synthesized in various dimensions for example, in one dimensional geometry: LiCuSbO$_4$ \cite{dutton2012quantum}, LiCuVO$_4$ \cite{mourigal2012evidence} and Rb$_2$Cu$_2$Mo$_3$O$_{12}$ \cite{hase2004magnetic,yagi2017nmr}; in ladder like  geometry: SrCu$_2$O$_3$ \cite{sandvik1996spin} and (VO)$_2$P$_2$O$_7$ \cite{johnston1987magnetic,dagotto1996surprises}; in two dimensional systems such as  ZnCu$_3$(OH)$_6$Cl$_2$(Herbertsmithite) \cite{shores2005structurally,helton2007spin,han2012fractionalized} and $\alpha$-RuCl$_3$ \cite{banerjee2016proximate}. Their properties have been rigorously explored theoretically as well \cite{broholm2020quantum,lake2005quantum,norman2016colloquium,takagi2019concept}.
In presence of a finite magnetic field, these systems exhibit plethora of quantum phases like multipolar \cite{fouet2006condensation}, spin nematic \cite{shannon2006nematic}, vector chiral \cite{hikihara2008vector}, plateaus state at various fractional 
magnetization \cite{leonov2015multiply,lee2008end,balents2010spin, momoi2000magnetization, schulenburg2002macroscopic}. 
 To explain the physical properties of these systems various models have  been proposed like Heisenberg spin-1/2  $J_1$-$J_2$ model  for one dimensional (1D) spin chain \cite{majumdar1969next,white1994resonating,chitra1995density,
 soos2016numerical}, Heisenberg anti-ferromagnetic spin-1/2  model on Shastry-Sutherland lattice (SSL) \cite{shastry1981exact} and  square lattice \cite{chakravarty1989two} , $J_1$-$J_2$ model on square lattice \cite{dagotto1989phase,sirker2006j} etc.

In general, frustrated 2D magnets are found in layered materials \cite{kundu2020gapless} 
and these 2D geometries: square \cite{manousakis1991spin}, triangular \cite{shirata2012experimental}, 
kagome \cite{helton2007spin} and SSL \cite{kageyama1999exact,zayed20174,lee2019signatures} are  particularly 
interesting because 2D is supposed to be a critical dimension in the purview of Mermin-Wagner 
theorem \cite{mermin1966absence}. We are particularly interested in SSL which is shown in Fig. \ref{ssl_figure}. 
This structure is similar to a square lattice except alternate squares have a diagonal bond. If a isotropic 
Heisenberg spin-1/2 model on the SSL is considered with antiferromagnetic exchange along the square
$J_x=J_y$ and diagonal exchange $J=2J_x=2J_y$ then the model can be solved exactly \cite{shastry1981exact}. 
The ground state of this system has N\'eel and dimer order for small and large $J/J_y$ limit respectively considering 
($J_x=J_y$).  Corboz \textit{ et al.} \cite{corboz2013tensor} predicted 
the existence of a plaquette phase between N\'eel and dimer region using the PEPS technique, whereas, very recently, 
Yang \textit{et al.}\cite{yang2021quantum} predicted a spin liquid phase. While it's exact nature has been controversial, 
Ronquillo and Peterson \cite{ronquillo2014identifying} predicted the topological gs.

Among the layered magnetic materials, spin-1/2 layered copper oxyhalides (CuX)A$_{n-1}$B$_n$O$_{3n+1}$ are frustrated 
2D magnets which have many interesting features just by tuning the composition \cite{}. 
The CuX layers are sandwiched by non-magnetic layers, the anion orbitals are  involved 
in exchange pathways but the cation orbitals replacement keeps the magnetic layer 
homogeneous. However, a small change in exchange interaction can be tuned by changing the lattice parameters, 
electrostatic fields, and crystal-field splittings \cite{tsirlin2012cucl}. 
The tuning of anion and cation composition can drive the system across the  quantum critical points. 
Experimental results suggest that the gs can be tuned to have a collective singlet  with a spin 
gap in (CuCl)LaNb$_2$O$_7$, \cite{kageyama2005spin,kageyama2005anomalous,kitada2007bose,yoshida2007magnetic}, 
a collinear stripe magnetic order in (CuBr)LaNb$_2$O$_7$, \cite{oba2006collinear} and a 
magnetization plateau at 1/3 of the saturated moment in (CuBr)Sr$_2$Nb$_3$0$_{10}$ \cite{tsujimoto20071}. 

(CuCl)LaNb$_2$O$_7$ was initially predicted as $S=1/2$ frustrated square lattice \cite{kageyama2005spin}, however, 
band structure calculations revealed that the simplest model for this material can be best described 
as  strong antiferromagnetic (AFM) exchange interaction between fourth neighbours forming a strong singlet dimer 
and these dimers are coupled together by ferromagnetic (FM) interactions i.e this model looks like the 
Shastry-Sutherland model with ferromagnetic exchange $J_x$ and $J_y$ along square and 
antiferromagnetic diagonal exchange $J$ \cite{tassel2010ferromagnetically}. 
This model was theoretically studied using  mean-field and exact diagonalization (ED) methods 
on a small cluster  and they predicated a plethora of phases in exchange parameter space 
\cite{furukawa2011ferromagnetically}. In the intermediate FM coupling limit, they reported 
two types of stripe phases, $(0,\pi)$ and $(\pi,0)$,  separated by a non-collinear spiral phase. 
The dimer singlet stabilizes for $J_x,J_y < J/2$, whereas ferromagnetic gs is stable 
for large  $J_x$ and $ J_y$ \cite{furukawa2011ferromagnetically}. We notice that the phase 
boundaries as well as the existence of different phases calculated from ED are not consistent with that 
calculated from  mean-field, for example, non-collinear phase does not appear in exact diagonalization results, 
but is present in the mean field calculations. Therefore,  it is very intriguing 
to explore such an interesting model with a more sophisticated numerical tool like density 
matrix renormalization group (DMRG) method which can give accurate results for large lattice sizes.


In this work, we explore the ferromagnetically coupled $S=1/2$ dimers on SSL with the DMRG method 
and re-investigate the quantum phase diagram of this model. The phase diagram is based on the nature  
of spin-spin correlations and ground state energy variation. The gs exhibits predominantly six 
types of phases: two types of stripe order with wave vector $(0,\pi)$ and $(\pi,0)$ for large 
value $J_y$ and $J_x$ respectively.  A perfect dimer phase exists for $J_x=J_y$  and this phase 
separates two types of spiral phases namely $X$-spiral with wave vector $(\theta,0)$ 
and $Y$-spiral with wavevector $(0,\theta)$, where $\theta$ is the pitch angle. 
In the large limit of $J_x$ and $J_y$ the gs has ferromagnetic behaviour. We also explore the effect of $J_x$ and $J_y$ 
on spiral behaviour and pitch angle and we notice that spin ordering is very short range in most of the parameter regimes.   

The paper is organized as follows. In Sec. \ref{sec-II} , the model of 
the ferromagnetically coupled SSL and numerical methods  are discussed.  In Sec. \ref{sec:result}, all the 
numerical results are presented and this section has four subsections. The quantum phase diagram is presented in 
 Sec. \ref{sec-IIIA}. Sec. \ref{sec-IIIB} and Sec. \ref{sec-IIIC} discuss gs energy 
 and spin-spin correlation in various phases of the phase diagram. The pitch angles are discussed in  Sec. \ref{sec-IIID}. 
 Results are discussed and compared with literature in  Sec. \ref{sec-summary}. In an appendix we presented results for the ground state energy per site and spin-spin correlation for various bond dimensions ($m$) and various system sizes.

\section{ Model Hamiltonian and numerical methods \label{sec-II}}

\begin{figure}[h]
\includegraphics[width=\linewidth]{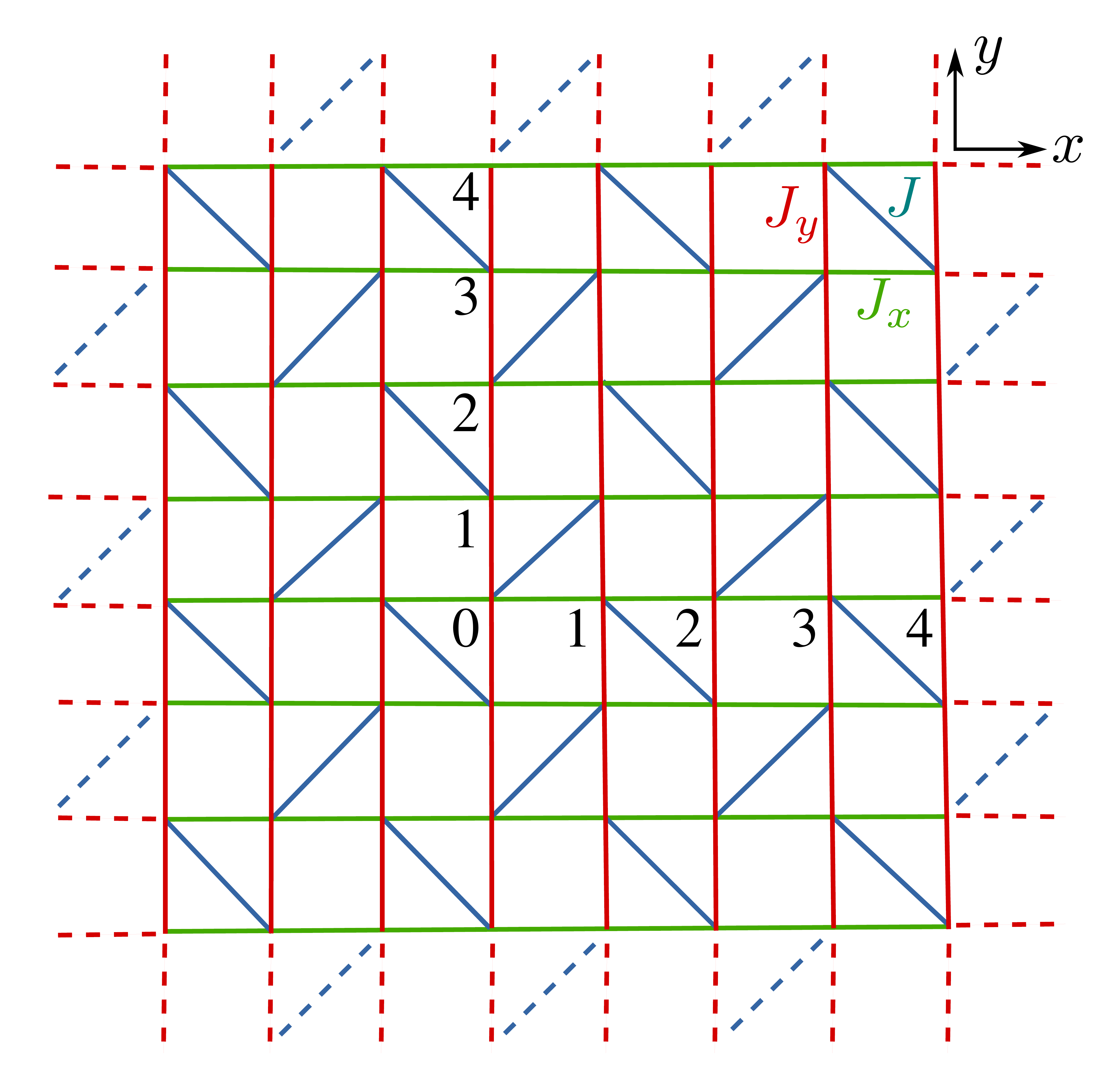}
\caption{Figure shows a schematic diagram of SSL.
The coupling constants along $x$-axis ($J_x$) and along $y$-axis ($J_y$) are FM and the diagonal bond 
couplings $J$ are AFM in nature. The site index $0$ is the reference site and the numbers along 
$x$-axis and $y$-axis represent the distance ($r$) from the reference site.} 
\label{ssl_figure}
\end{figure}

We consider a Heisenberg spin $S=1/2$ model on SSL where only diagonal interaction $J=1$ 
is antiferromagnet and sets the energy scale of the system.  The strength of ferromagnetic exchange interaction along the 
$x$-axis and $y$-axis on the square is represented by $J_x$ and $J_y$ respectively. The arrangement of the exchange interactions are shown in Fig. \ref{ssl_figure}. Now onward we will call this model as Shastry-Sutherland (SSM) and can be written as
\begin{equation}
H= -J_x\sum_{<ij>_x} \vec{S}_i \cdot \vec{S}_j - J_y\sum_{<ij>_y} \vec{S}_i \cdot \vec{S}_j + J\sum_{<ij>_d} \vec{S}_i \cdot \vec{S}_j
\label{ssm_hamiltonian}
\end{equation}
where the first sum runs for NN bonds along the $x$-direction, the second sum runs for NN bonds along the 
$y$-direction and the last sum runs for NN diagonal bonds in the square. 

We use the exact diagonalization for system size up to 32 sites  and  the density matrix renormalization group (DMRG) method 
~\cite{white1992density,white1993density,schollwock2005density,hallberg2006new} for large system size. 
The DMRG method is a state of art numerical method 
to handle the large degrees of freedom for a many-body Hamiltonian in low dimensions. 
This method is based on the systematic truncation of irrelevant degrees of freedom 
while growing the system sizes and optimising the wavefunction while doing the finite DMRG algorithm. 
In this work, we use a modified DMRG method \cite{kumar2010modified} 
in which 4-new sites are added at each step to reduce the number of times of renormalization of 
operators used to build the super block \cite{kumar2010modified}. We retain up to $m=900$ block 
states which are the eigenvectors of the system block density matrix corresponding to 
the largest eigenvalues. The chosen value of `$m$' keeps the truncation error to less than 
$\sim 10^{-8}$. We also carry out $10 -12$ finite sweeps for improved convergence and 
to optimise the wave function. We use cylindrical geometry of the SSL and a periodic boundary is applied 
along the width of the lattice and open boundary along the length. The largest system size studied is up 
to  $12 \times 8$ (length $\times$ width) system size. We have analysed the convergence of energy 
with the various values of $m$ in appendix. The per-site energies for $12\times 4$  
system are shown for $m=256$, 512 and 900, whereas for $12\times 8$ system, it is shown for  $m=256$, 512 and 700
in table \ref{tab:energy}.  We notice that $m=512$ is sufficient for accuracy up to $5^{th}$ decimal place for $12\times 4$  
and $4^{th}$ for $12\times 8$ system. We also show the dependence of the spin-spin correlation function on $m$ 
in table \ref{tab:spin-spin_48} and \ref{tab:spin-spin_96} of appendix. We notice that the spin-spin correlations 
are accurate up to $4^{th}$ decimal places. Interestingly this model gives amazing accuracy of energies as well 
as correlation function and these accuracy may be attributed to the short-range correlation length which 
is approximately $~4$ lattice unit.     

\section{Results \label{sec:result}}
\begin{figure}[]
\includegraphics[width=\linewidth]{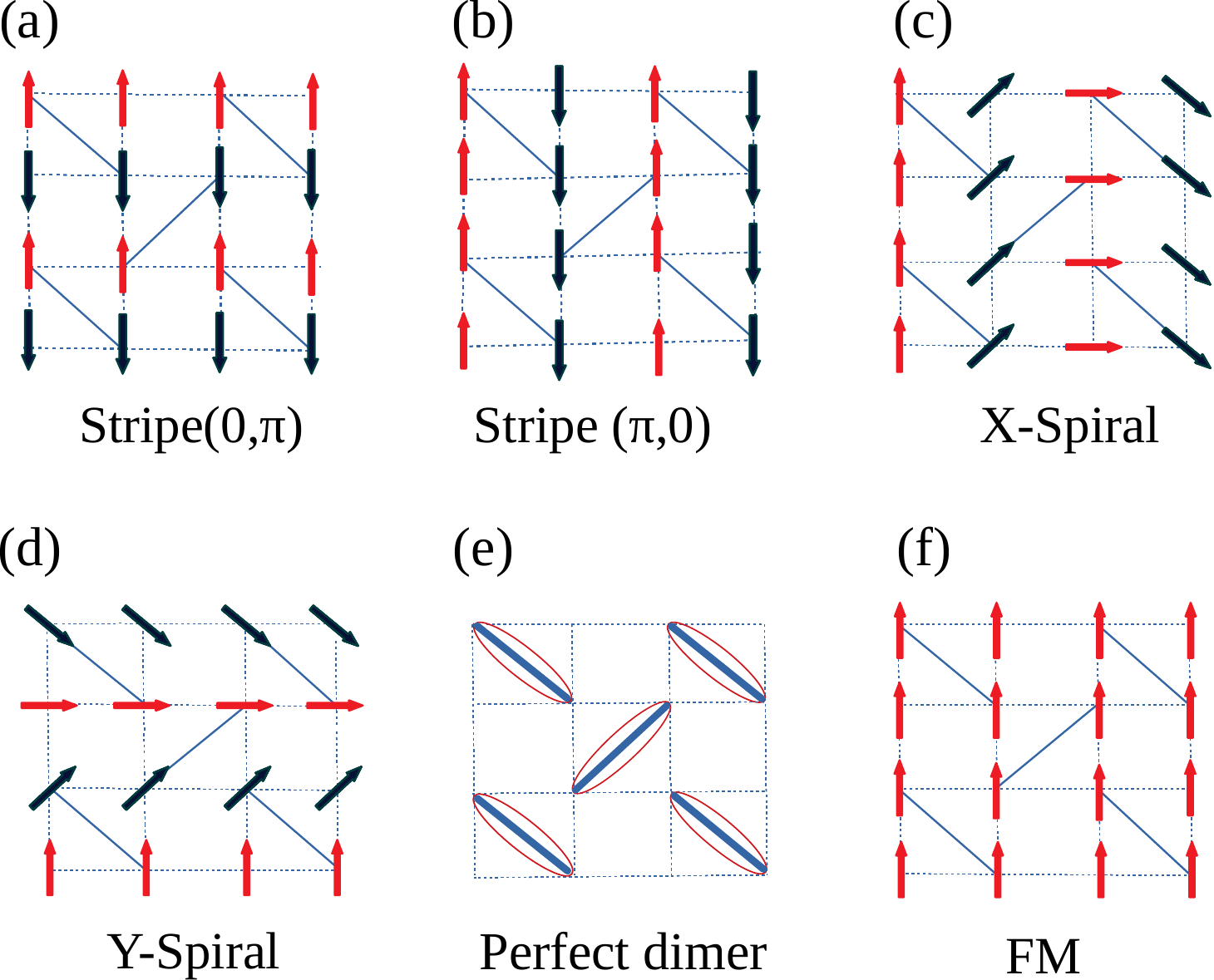}
\caption{Figure shows a schematic diagram of different phases present in the 
ferromagnetically coupled dimers on  the SSL, where arrows represent the spin alignments and 
the ellipses are the dimers (defined in the main text). 
(a) and (b) are two stripe phases with ordering wave vector $(0,\pi)$ and ($\pi,0$) 
respectively. (c) and (d) are two non-collinear spiral phases, the $X$-spiral and 
$Y$-spiral. (e) and (f) are the perfect dimer and FM phase respectively.}
\label{schematic_dia}
\end{figure}

To identify the existence of various phases of the model with isotropic Heisenberg exchange in 
Eq. \ref{ssm_hamiltonian}, we calculate spin-spin correlation $C(r)$ and gs energies. 
 There are six different phases in this model: 
(i) stripe $(0,\pi)$ where the spin configuration has a propagation vector $(k_x,k_y)=(0,\pi)$ 
and the system forms stripes along the $x$-direction where all the spins along $x$-direction are arranged 
ferromagnetically while spin modulates with  wavevector $\pi$ along the $y$-direction as shown in Fig. \ref{schematic_dia}(a).
(ii) In the stripe $(\pi,0)$ phase, stripes run along the $y$-direction where spins are arranged ferromagnetically in 
$y$-direction while spin wave has  wavevector $\pi$ along the $x$-direction as shown in Fig. \ref{schematic_dia}(b).
(iii) The ground state with a $X$-spiral phase has a non-collinear arrangement of spins along the
$x$-direction and ferromagnetic ordering along the $y$-direction as shown in Fig. \ref{schematic_dia}(c). This phase appears in case of highly frustrated model with AFM 
$J_x=J_y$ in Ref. [\onlinecite{shastry1981exact}].
(iv) In $Y$-spiral phase, spins have a non-collinear arrangements along the
$y$-direction and ferromagnetic ordering sets in along $x$-direction as in Fig. \ref{schematic_dia}(d). 
(v) In the perfect dimer phase the gs wave function can be represented as a
product of dimer singlets: 
$|\psi\rangle\equiv \prod_d ~\frac{1}{\sqrt{2}}(|\uparrow\downarrow\rangle - |\downarrow\uparrow\rangle)_d$, 
where $d$ labels a dimer \cite{shastry1981exact,furukawa2011ferromagnetically}. In this phase, the 
dimers are formed along the diagonal bonds as shown in Fig. \ref{schematic_dia} (e). 
The spin-spin correlation $C(r)=-0.75$ along the diagonal bonds and  $C(r)=0$ along any other bonds on the square. 
(vi) When the FM couplings dominate over the diagonal antiferromagnetic exchange interaction the FM 
phase appears. In this phase, all spins are aligned in the same direction as shown in Fig. \ref{schematic_dia}(f). 
\subsection{Quantum Phase Diagram \label{sec-IIIA}}
\begin{figure}[h!]
\includegraphics[width=\linewidth]{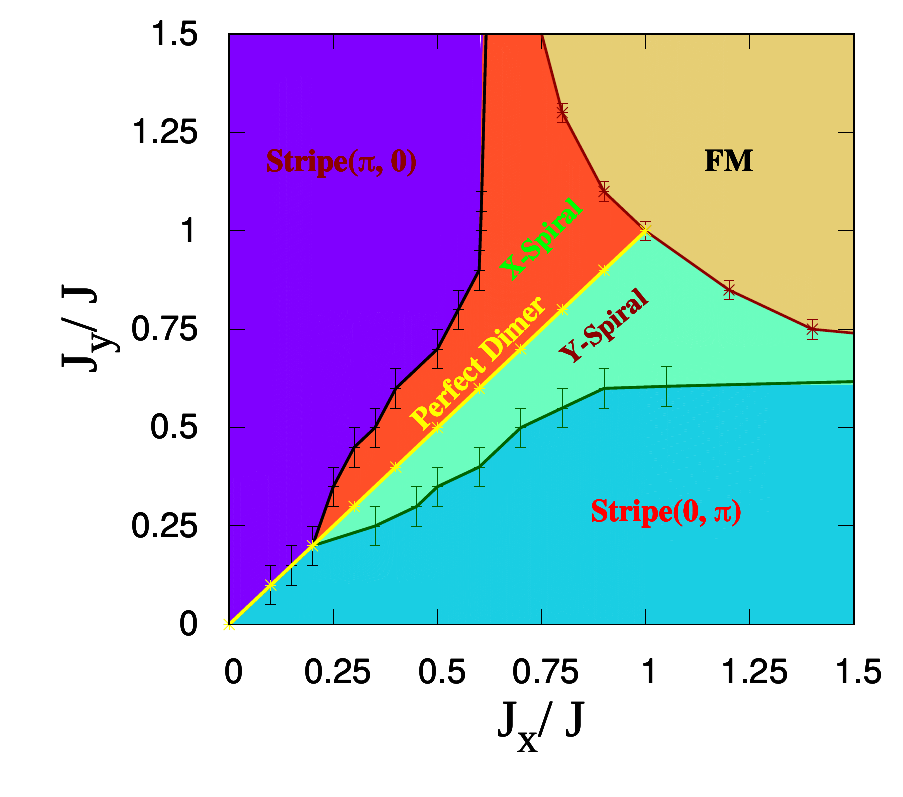}
\caption{Figure shows the quantum phase diagram of the SSM. The phase 
diagram comprises six phases: Two stripe order phases with ordering wave vector $(0,\pi)$ and 
$(\pi,0)$, two spiral orders along both the $x$- and $y$-directions, the $X$-spiral and $Y$-spiral 
phase respectively. The other two phases are a perfect dimer phase (the yellow line separating two 
spiral phases) and a phase with FM spin ordering along all directions of the lattice.}
\label{ssm_phasediagram}
\end{figure}

In this section, the quantum phase diagram is presented  in $J_x$ and $J_y$ parameter 
space as shown in Fig.\ref{ssm_phasediagram} for a given diagonal exchange $J=1$.
For a large value of $J_x$ or $J_y$ stripe phases with wave vectors  $(0,\pi)$ and  $(\pi,0)$ stabilized in the 
gs of the system. In this region, frustration is considerably small due to the small or large value of 
$J_x/J_y$. The strong FM coupling bonds prefer FM spin orderings, whereas, weak FM 
coupling bonds  prefer antiferromagnetic spin ordering to satisfy the diagonal AFM exchange.
For the moderate exchange strength $J_x <1 $ and $J_y < 1$ the gs has non-collinear spin ordering. 
Depending on the relative exchange strength of $J_x$ and $J_y$ there 
are two types of spiral phases: $X$-spiral and $Y$-spiral. These two spiral phases are 
 separated by a perfect dimer phase formed along the line $J_x=J_y$ and extended up to 
$J_x, J_y \lesssim J$. The perfect dimer region  is confined along the 
$J_x=J_y$ line as $C(r=2)$ continuously increases with 
$|J_x-J_y|$. In the strong coupling region of $J_x$ and $J_y$, 
long-range FM magnetic ordering sets in. 

\subsection{Ground state energy \label{sec-IIIB}}
\begin{figure}[h]
\includegraphics[width=\linewidth]{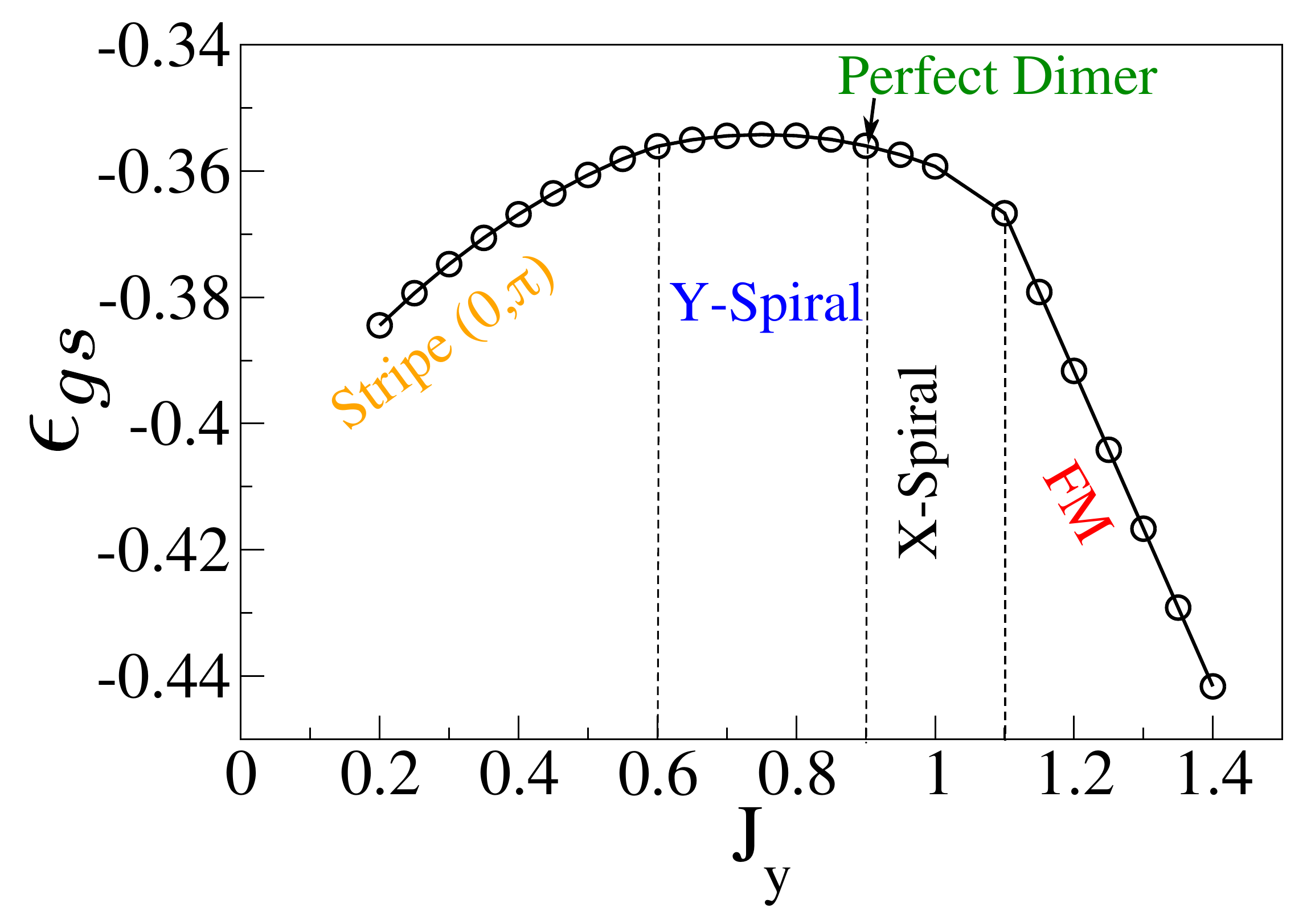}
\caption{Figure shows gs energy per site ($\epsilon_{gs}$) versus $J_y$ plot at $J_x=0.9$ for $12\times 4$ lattice sites of the SSM.
Around $J_y=1.1$ we show a kink in the energy curve indicating a sharp phase transition from $X$-spiral to FM phase. The transition between the two spiral phases is not sharp. Around $J_y=0.6$ system transits from $Y$-spiral to stripe $(0,\pi)$ phase. (Shaded lines are only for eye guide of the transition points).}
\label{ssm_energy}
\end{figure}

Ground state energies are calculated for the model in Eq. \ref{ssm_hamiltonian} on a cylindrical geometry 
with periodic boundary condition (PBC) along the width ($y$-axis) and open boundary 
condition along the length ($x$-axis), and $12\times 4$ system is used to 
calculate the accurate gs. In Fig. \ref{ssm_energy} the gs energy per site
$\epsilon_{gs}$ is 
plotted as a function of $J_y$ by keeping $J_x$ fixed value of $J_x=0.9$. Around $J_y\sim 1.1$ 
we saw a discontinuity in the gs energy, indicating a first-order phase transition from FM 
to $X$-spiral phase. The maxima of the energy curve is close to $J_x=J_y \sim 0.9$ 
which indicates the dimer line, and the small deviation of maxima from dimer line 
$J_x=J_y$ is due to the rectangular geometry of the lattice. The maxima shifts to 
$J_x=J_y$ line as we increase the width of the lattice. 
The smooth variation of $\epsilon_{gs}$ with $J_y$
indicates the second-order dimer transition. No signature of spiral and stripe 
transitions is detected from the gs energy variation.  
\subsection{Spin-spin correlation \label{sec-IIIC}}
In this subsection, we present spin-spin correlation $C(r)$ for various parameter regimes 
to validate the  existence of different phases in the phase diagram in Fig. \ref{ssm_phasediagram}. 
We are dealing with isotropic system and total spin-spin correlations $C(r)= \langle \vec{S}_i \cdot \vec{S}_{i+r}\rangle= 3 \langle S^z_i S^z_{i+r} \rangle$ where $S^z_i$ represents the $z$-component of the spin operator at 
the reference site $i$. $r$ is the distance between the spin site and reference site along the $x$- and $y$-axis as shown in Fig. \ref{ssl_figure}. 
The spin Hamiltonian in Eq. \ref{ssm_hamiltonian} has $SU(2)$ symmetry and  one expects equal spin-spin correlation 
for all three spin components in the singlet sector.
To understand the spin arrangements in different phases we have calculated $C(r)$ along different spatial directions. In Fig. \ref{ssl_figure}, the site 
with $0$ index represents the reference site, from which  correlations are calculated in various directions, 
and distances $r$ are shown in various directions with numerials as shown in Fig. \ref{ssl_figure}.
\begin{figure}[h]
\includegraphics[width=\linewidth]{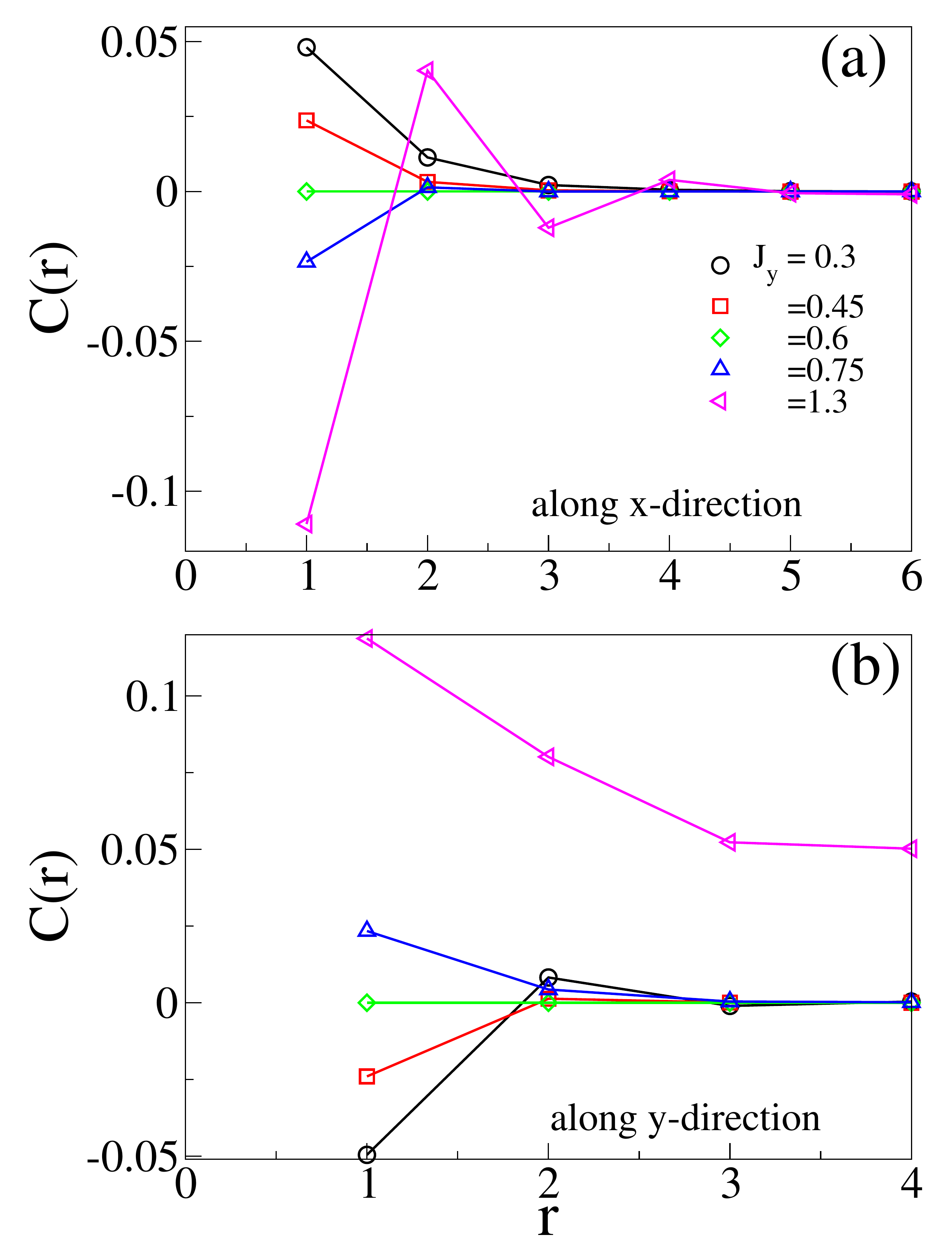}
\caption{Figure shows the variation of spin-spin correlations $C(r)$ with distance $(r)$ for $12\times 8$ system size. Spin-spin correlations are calculated for different $J_y$ values by keeping $J_x$ fixed at $0.6$. (a) $C(r)$ are calculated along $x$-direction. (b) $C(r)$ are calculated along $y$-direction. 
}
\label{ssm_corr_48leg}
\end{figure} 
In Fig. \ref{ssm_corr_48leg}(a) we presented 
$C(r)$ along the $x$-direction, whereas Fig. \ref{ssm_corr_48leg} (b) represents the same for $y$-direction. 
In all plots, we have fixed $J_x=0.6$ and calculated $C(r)$ for different $J_y=0.3$, 0.45, 0.6, 0.75 and 
1.3, where $J_y$ values correspond to different phases in the phase diagram.
 
In Fig. \ref{ssm_corr_48leg}(a) for $J_y=0.3$, all the values of $C(r)$ 
are positive which correspond to an FM spin arrangment along $x$-direction. 
Whereas in Fig. \ref{ssm_corr_48leg}(b) for same value of $J_y=0.3$, $C(r)$ shows 
an antiferromagnetic arrangement of spins along $y$-direction. This kind of spin arrangements 
correspond to a stripe spin ordering with  a wave vector $(0,\pi)$ (see Fig.\ref{schematic_dia}(a)). 
For $J_y=1.3$ the behaviour of the spin correlation along the $x$-direction interchanges 
with the behaviour along $y$-direction i.e antiferromagnetic arrangement along the $x$-direction 
and ferromagnetic arrangement along the $y$-direction. This represents another kind of stripe phase with 
wave vector at $(\pi,0)$ as shown in Fig. \ref{schematic_dia}(b).

At $J_y=0.45$, $C(r)$ along $x$-direction are all positive and correspond to an FM spin alignment 
in this direction, whereas in the $y$-direction the $C(r)$ is non-collinear i.e pitch angle
is different from 0 or $\pi$. The non-collinear spin arrangement is shown along the $y$-direction and 
the $Y$-spiral phase can be seen in Fig. \ref{schematic_dia}(d). For $J_y=0.75$, $C(r)$ shows a 
non-collinear behaviour along $x$-direction whereas ferromagnetic behaviour along the $y$-direction. 
The schematic of $X $-spiral phase is shown in Fig. \ref{schematic_dia}(c). 

When $J_x=J_y (=0.6)$, $C(r)$ along $x$-direction and $y$-direction are exactly zero, 
whereas, we found the correlation for diagonal bonds are $-0.75$ 
This shows a perfect dimer formation along the diagonal bonds and this phase 
is shown in Fig. \ref{schematic_dia}(e). A perfect dimer phase is formed along $J_x=J_y \lesssim 1$ line. In low $J_x$ and $J_y$ limit  $C(r>1)$ have non zero value for  $|J_x-J_y|>0$, therefore the perfect dimer phase is only restricted to $J_x=J_y$.
In the FM phase region $C(r)$ have all positive values in all directions. 

\subsection{Pitch angle \label{sec-IIID}}
\begin{figure}[h]
\includegraphics[width=\linewidth]{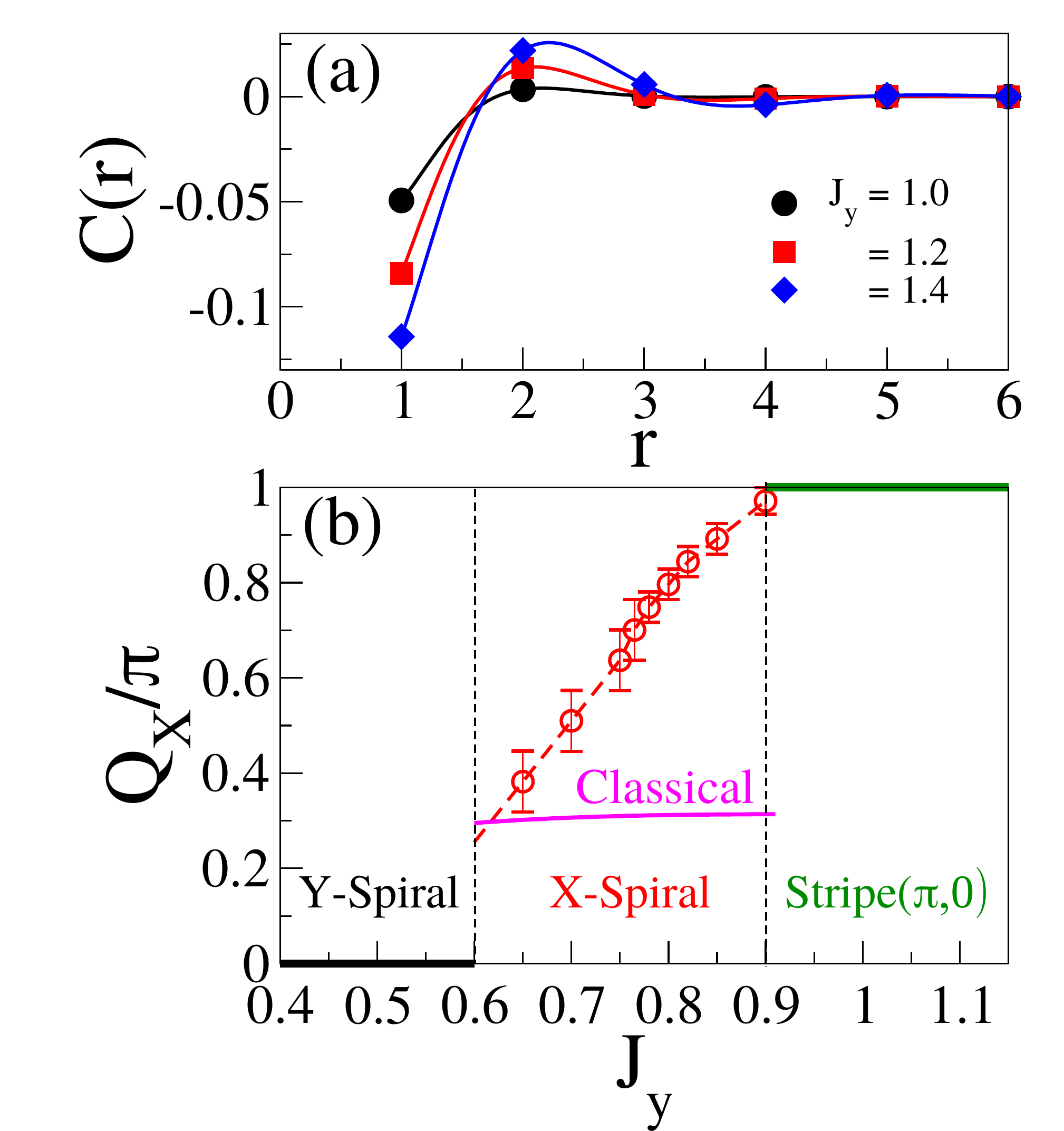}
\caption{Figure (a) shows the spin-spin correlation $C(r)$ along length ($x$-direction) for different $J_y$ at $J_x=0.7$. The solid curves represent respective fits with exponential decaying function.  In figure (b) we have plotted pitch angle $(Q_x/\pi)$ as a function of $J_y$ at $J_x=0.6$ 
for different phases of the model, where $Q_x$ is the pitch angle for the $X$-spiral phase varying along $x$-direction. The pink line is the value of pitch angles calculated using Eq. \ref{eqn:spiral}.  In both plots we have used $12\times 4$ system size.}
\label{ssm_pitchangle_fit}
\end{figure}
 In a geometrically frustrated system wavevector or the pitch angle of a non-collinear spin ordering, in general, depends 
on the competing exchange interactions \cite{furukawa2012ground,parvej2017multipolar,maiti2019quantum} and therefore, 
it is important to understand behaviour of pitch angle $\theta$ in various exchange interaction limits to 
quantify spin modulation in terms of the pitch angle $Q_x$ and $Q_y$. 
{Fig. \ref{ssm_pitchangle_fit}(a) shows spin-spin correlation  $C(r)$ for $J_x=0.7$ for different $J_y$ values inside the $X$-spiral region. $C(r)$ shows the exponentially decaying behaviour with distance r. It indicates the non-collinear spin ordering along $x$-direction, representing the $X$-spiral phase. $C(r)$ in Fig. \ref{ssm_pitchangle_fit}(a) shows an exponential decay, representing a very short-range correlation in the spiral region.} 
Pitch angle $(Q_x)$ can be calculated in the $X$-spiral phase, by fitting $C(r)$ with the following equation
\begin{equation}
C(r)=a_0\cos(Q_x r + c)~e^{- \frac{r}{\xi}}~,
\label{Qx_fiting}
\end{equation}
where $\xi$ is the correlation length, $a_0$ and $c$ are  constants. To calculate the pitch angles in 
the $Y$-spiral phase, a similar correlation function can be applied as above, where $Q_x$ will be replaced by $Q_y$.
The variation of pitch angle $Q_x$ with $J_y$ is presented in Fig. \ref{ssm_pitchangle_fit}(b) and we notice 
that $Q_x$ varies between $0$ to $\pi$. We also notice that $Q_x$ changes sharply near $J_x\sim 0.6$. 
$Q_x$ and $Q_y$ variation with $J_y$ and $J_x$ for a classical 
system \cite{furukawa2011ferromagnetically} can be given as 
\begin{eqnarray}
\cos(Q_x) &=& -\frac{J_y}{J}+\frac{J}{4J_x}\left(\frac{J_y}{J_x}-\frac{J_x}{J_y}\right),\nonumber\\
	\cos(Q_y) &=& -\frac{J_x}{J}+\frac{J}{4J_y}\left(\frac{J_x}{J_y}-\frac{J_y}{J_x}\right)~.
\label{eqn:spiral}
\end{eqnarray}
The classical results seem to deviate significantly 
from our calculated values. The DMRG results are shown as dotted line with circles 
whereas the classical results are shown as solid line as shown in Fig. \ref{ssm_pitchangle_fit}(b). 
The calculated values of $Q_x$ suggest that it is zero below the dimer line and increase rapidly with 
$J_y$ and reaches to $\pi$ for $J_y\sim 1.0$. In the stripe $(\pi,0)$ phase the $Q_x$ is $\pi$, whereas it is zero in 
the stripe $(0,\pi)$ phase. These systems have correlation length $\xi  \approx 4 $ lattice units, therefore, 
there is an error in calculating the $Q_x$ and it is represented by error bars. The system remains invariant by 
exchanging the $J_y$ to $J_x$ interaction, therefore $Q_x-J_y$  and $Q_y-J_x$ curves 
have the same nature as shown in $Q_x-J_y$ curve.  

\section{Summary and conclusion \label{sec-summary}}
In this work, we construct a new quantum phase diagram of ferromagnetically coupled $S=1/2$ on the SSL.
Exchange couplings along width $J_y$ and length $J_x$ are ferromagnetic, whereas, the exchange couplings along 
diagonal bonds $J$  are antiferromagnetic. The quantum phase diagram of the SSM in Eq. \ref{ssm_hamiltonian} 
consists of six phases and the phase boundaries are calculated based on spin-spin correlation and the gs energies obtained 
using the ED and the DMRG methods. Our numerical calculations are done upto $12\times8$ lattices 
and we have used PBC along the width and OBC along length. Almost 
everywhere in the phase space the order is short range and correlation length $\xi$ is less than equal 
to  $4$  lattice unit, therefore, most of our calculations give reliable results for the two dimensional 
lattice of this model. The six phases in the phase diagrams are: (I) stripe $(0,\pi)$, 
(II) stripe $(\pi, 0)$, (III) perfect dimer, (IV) $X$-spiral, (V) $Y$-spiral and (VI) ferromagnetic state. 
Our analysis also confirms the existence of spiral phases in this model for moderate FM couplings strength. 

Our DMRG results are very different from the Schwinger-Boson mean-field theory 
in Ref. [\onlinecite{furukawa2011ferromagnetically}]. Although quantum phases predicted by mean-field theory are 
also found in the DMRG results but phases boundaries are quite different. DMRG calculations suggest 
that perfect dimer singlet phase is confined to only on $J_x=J_y$ line, whereas the mean-field 
calculation suggests a large area of this regime. Our result is consistent with the ED 
results \cite{furukawa2011ferromagnetically}. Our results also suggest that second nearest neighbour 
correlation increases continuously  with $|J_x-J_y|$ i.e. short range but finite correlation exists
in the neighbourhood of dimer line except the $J_x=J_y$ line where only diagonal correlations 
are non zero. The mean-field results suggest
the non-collinear spin wave along the $x$- and $y$-direction but ED does not confirm the results \cite{furukawa2011ferromagnetically}. DMRG 
results confirm both types of the non-collinear phases. 
The mean-field pitch angle variation is small compared to the DMRG value in most parts of the parameter regime. 

 Tassel \textit{et al.} suggested that (CuCl)LaNb$_2$O$_7$  has ferromagnetic $J_y/J \sim 0.39$ 
 and $J_x/J \sim 0.38$ \cite{tassel2010ferromagnetically} and therefore, we expect this system 
 should behave like dimer as it is on $J_x=J_y$ dimer singlet line. Although, 
 Tsirlin \textit{et al.} shows different exchange interactions using the two types of DFT 
 calculations \cite{tsirlin2012cucl}. However, the data of inelastic neutron scattering (INS) on powder 
 samples shows that the dynamical structure factor has $S(Q,\omega)$ maxima around $Q\sim 0.5$\AA ,  
 i.e. it is in a non-collinear regime. In our opinion this  material is in the neighbourhood of 
 dimer phase but detailed theoretical investigation is required to understand the spin configuration in the system.   

In conclusion, we numerically studied the SSM and constructed a new quantum 
phase diagram using the DMRG method. We have also calculated 
the correlation function and pitch angle which can be directly connected to INS data. The phase boundaries 
calculated from DMRG results are different from that of mean-field and ED calculations \cite{}. We hope that
interesting quantum phases in real material like (CuCl)LaNb$_2$O$_7$ and 
others \cite{tassel2010ferromagnetically,uemura2009muon,kitada2009quantum,tsujimoto2009synthesis} can be 
revisited in light of this study. The short-range spiral phase in the frustrated regime of the model 
can be manipulated by external probe like a magnetic field, doping, etc. and might lead to 
many interesting phases. The effect of  field on  phase diagram of the model in Eq. \ref{ssm_hamiltonian} is 
 still an open problem.   

\section*{Acknowledgments}
MK thanks SERB for financial support through grant sanction number CRG/2020/000754. MC thanks DST-INSPIRE for financial support and also thanks S K Saha for fruitful discussions.


%

\section{Appendix \label{appendix}}
\begin{table*}[h]
\normalsize
\begin{center}
\begin{tabular}{|C{1.8cm}|C{1.8cm}|C{1.8cm}|C{1.8cm}|C{3cm}|}

\hline
$L_x\times L_y$ & $J_x$ &  $J_y$ &  $m$ & $\epsilon_{gs}$   \\ 
\hline
\hline
 \multirow{3}{*}{$12\times 4$}& \multirow{3}{*}{0.6} & \multirow{3}{*}{0.21} & 256 & -0.357791 \\ 
\cline{4-5}
 &  & & 512 & -0.357795 \\ 
\cline{4-5}
 &   & & 900 & -0.357798 \\ 
\hline
\multirow{3}{*}{$12\times 8$}& \multirow{3}{*}{0.6} & \multirow{3}{*}{1.0} & 256 & -0.36989 \\ 
\cline{4-5}
 &  &  & 512 & -0.36987 \\  
\cline{4-5}
 & & & 700 & -0.36984 \\
\hline
\end{tabular}
\end{center}
\caption{\label{tab:energy} 
The table shows the gs energy per site ($\epsilon_{gs}$) calculated for various bond dimensions ($m$). For the system size $12\times 4$ gs energy calculated for $J_x=0.6$ and $J_y=0.21$, whereas for $12\times 8$ system size, it is calculated for $J_x=0.6$ and $J_y=1.0$.}
\end{table*}

\begin{table*}[h]
\normalsize
\begin{center}
\begin{tabular}{|C{1.8cm}|C{1.8cm}|C{3cm}|C{3cm}|C{3cm}|}

\hline
direction & $r$ & $m=256$ &  $m=512$ &  $m=900$    \\ 
\hline
\hline
 \multirow{5}{*}{length} & 1 & 0.02154 & 0.02153 & 0.02152 \\ 
\cline{2-5}
 & 2 & 0.00652 & 0.00651 & 0.00650 \\ 
\cline{2-5}
 & 3 & 0.00182 & 0.00182 & 0.00178 \\  
\cline{2-5}
 & 4 & 0.00056 & 0.00056 & 0.00053 \\
\cline{2-5}
 & 5 & 0.00015 & 0.00015 & 0.00014 \\
\hline
\multirow{2}{*}{width} & 1 & -0.02269 & -0.02269 & -0.02267 \\ 
\cline{2-5}
 & 2 & 0.01006 & 0.01006 & 0.01008 \\ 
\hline
\end{tabular}
\end{center}
\caption{\label{tab:spin-spin_48} 
The table shows spin-spin correlation $\langle S^z_i S^z_{i+r}\rangle$ calculated for various bond dimensions ($m$) for $12\times 4$ system size. The correlation calculated along length and width for $J_x=0.6$ and $J_y=0.21$, where the last three columns are the correlation values for the corresponding $m$.}
\end{table*}
\begin{table*}[h]
\normalsize
\begin{center}
\begin{tabular}{|C{1.8cm}|C{1.8cm}|C{3cm}|C{3cm}|C{3cm}|}

\hline
direction & $r$ & $m=256$ &  $m=512$ &  $m=700$   \\ 
\hline
\hline
 \multirow{6}{*}{length} & 1 & -0.02095 & -0.02097 & -0.02088 \\ 
\cline{2-5}
 & 2 & 0.00354 & 0.00338 & 0.00328 \\ 
\cline{2-5}
 & 3 & -0.00037 & -0.00032 & -0.00030 \\  
\cline{2-5}
 & 4 & 0.000058 & 0.000050 & 0.000042 \\
\cline{2-5}
 & 5 & -0.000006 & -0.000019 & -0.000004 \\
 \cline{2-5}
 & 6 & -0.000001 & 0.000033 & 0.000001 \\
\hline
\multirow{4}{*}{width} & 1 & 0.021093 & 0.021144 & 0.021042 \\ 
\cline{2-5}
 & 2 & 0.008490 & 0.008516 & 0.008512 \\ 
\cline{2-5}
 & 3 & 0.002574 & 0.002553 & 0.002526 \\ 
\cline{2-5}
 & 4 & 0.001771 & 0.001719 & 0.001769 \\ 
\hline
\end{tabular}
\end{center}
\caption{\label{tab:spin-spin_96} 
The table shows spin-spin correlation $\langle S^z_i S^z_{i+r}\rangle$ calculated for various bond dimensions ($m$) for $12\times 8$ system size. The correlation calculated along length and width for $J_x=0.6$ and $J_y=1.0$, where the last three columns are the correlation values for the corresponding $m$.}
\end{table*}
\end{document}